\documentclass[aps,pra,superscriptaddress,twocolumn,longbibliography]{revtex4-2}
\usepackage{units}
\usepackage{amsmath}
\usepackage{amsthm}
\usepackage{amssymb}
\usepackage{graphicx}
\usepackage{color}
\usepackage{xcolor}
\usepackage{bbold}
\usepackage[normalem]{ulem}

\usepackage[colorlinks=true, urlcolor=blue, citecolor=blue,linkcolor=blue,citebordercolor={1 0 0},linkbordercolor={0 0 1}]{hyperref}

\usepackage[linesnumbered, ruled,vlined]{algorithm2e}

\graphicspath{{./images/},{./imagesAppendix/}}

\theoremstyle{plain}

\theoremstyle{plain}

\ifx\proof\undefined
\newenvironment{proof}[1][\protect\proofname]{\par
	\normalfont\topsep6\p@\@plus6\p@\relax
	\trivlist
	\itemindent\parindent
	\item[\hskip\labelsep\scshape #1]\ignorespaces
}{%
	\endtrivlist\@endpefalse
}
\providecommand{\proofname}{Proof}
\fi
\theoremstyle{plain}

\theoremstyle{remark}

\newcommand{\bra}[1]{\langle #1|}
\newcommand{\ket}[1]{|#1 \rangle}
\makeatother
\newcommand{\braket}[2]{\langle #1 \vert #2 \rangle}

\newcommand{\idg}[1]{{\bfseries #1)}}

\renewcommand{\eqref}[1]{Eq.~(\ref{#1})} %

\newcommand\numberthis{\addtocounter{equation}{1}\tag{\theequation}}
\providecommand{\factname}{Fact}
\providecommand{\theoremname}{Theorem}
\providecommand{\claimname}{Claim}
\providecommand{\lemmaname}{Lemma}
\providecommand{\definitionname}{Definition}

\definecolor{KB}{rgb}{0.4,0.3,0.9}

\definecolor{THc}{rgb}{0.9,0.3,0.2}

\newcommand{\revA}[1]{{#1}}

\theoremstyle{definition}

\definecolor{adiyellow}{RGB}{200,150,0}    %

\newcommand{\subfigimg}[3][,]{%
	\setbox1=\hbox{\includegraphics[#1]{#3}}%
	\leavevmode\rlap{\usebox1}%
	\rlap{\hspace*{2pt}\raisebox{\dimexpr\ht1-0.5\baselineskip}{{\bfseries \large\textsf{#2}}}}%
	\phantom{\usebox1}%
}

\begin{document}

\title{Disentangling quantum autoencoder}

\author{Adithya Sireesh}
\email{asireesh@ed.ac.uk}
\affiliation{School of Informatics, University of Edinburgh, Scotland, United Kingdom}
\affiliation{Department of Computing,
Imperial College London SW7 2BX, UK}
\author{Abdulla Alhajri}
\affiliation{Quantum Research Center, Technology Innovation Institute, Abu Dhabi, UAE}
\author{M.~S. Kim}
\affiliation{Blackett Laboratory, Imperial College London SW7 2AZ, UK}

\author{Tobias Haug}
\thanks{Corresponding author}
\email{tobias.haug@u.nus.edu}
\affiliation{Quantum Research Center, Technology Innovation Institute, Abu Dhabi, UAE}
\affiliation{Blackett Laboratory, Imperial College London SW7 2AZ, UK}

\begin{abstract}
Entangled quantum states are highly sensitive to noise, which makes it difficult to transfer them over noisy quantum channels or to store them in quantum memory. 
Here, we propose the disentangling quantum autoencoder (DQAE) to encode entangled states into single-qubit product states. 
The DQAE provides an exponential improvement in the number of copies needed to transport entangled states across qubit-loss or leakage channels compared to unencoded states.
The DQAE can be trained in an unsupervised manner from entangled quantum data. For general states, we train via variational quantum algorithms based on gradient descent with purity-based cost functions, while stabilizer states can be trained via a Metropolis algorithm.
For particular classes of states, the number of training data needed to generalize is surprisingly low: For stabilizer states, DQAE generalizes by learning from a number of training data that scales linearly with the number of qubits, while only $1$ training sample is sufficient for states evolved with the transverse-field Ising Hamiltonian.
Our work provides practical applications for enhancing near-term quantum computers.
\end{abstract}

\maketitle

\section{Introduction}
Entanglement is a key feature of quantum mechanics that describes the non-local quantum correlations, and is a necessary ingredient for achieving quantum advantages in quantum communication, quantum sensing and quantum computing~\cite{horodecki2009quantum}.
However, at the same time, entanglement makes quantum states fragile, as the non-local correlations are highly susceptible to noise that affects quantum states in experiments. In particular, even noise that acts only on a local subsystem of the entangled state can already collapse the global wavefunction into a classical mixture of states without useful quantum correlations. 

To make states robust against noise, various methods have been proposed, such as quantum error correction, where the logical information is encoded redundantly into many physical qubits~\cite{lidar2013quantum}. However, quantum error correction is highly resource-intensive and not possible in noisy intermediate-scale quantum computers~\cite{bharti2021noisy}. Furthermore, error correction may be infeasible in scenarios where physical constraints exist, such as a limited number of qubits or long-distance communication between parties.

Alternatively, can reducing the entanglement of states enhance their robustness to noise?
In classical settings, reducing non-local correlations to simplify data has proven to be a fruitful approach. For example, in classical neural networks, variational autoencoders have been used to disentangle highly correlated information into factorized probability distributions~\cite{burgess2018understanding,mathieu2019disentangling}. This has allowed for finding simpler representations of complicated data that can enhance machine learning.

In quantum information, quantum autoencoders have been proposed to compress high-dimensional quantum states into a lower-dimensional subspace~\cite{wan2017quantum,romero2017quantum,lamata2018quantum}.
They can find a more compact representation of quantum states~\cite{bravo2021quantum,liu2024information,araujo2024schmidt,selvarajan2022dimensionality} to enhance information processing. 
Notably, quantum autoencoders can be used to denoise quantum states~\cite{bondarenko2020quantum,cao2021noise,mok2023rigorous}, and discover new quantum error correction codes~\cite{locher2023quantum}.
They have been applied to enhance quantum communication in photonic systems~\cite{zhang2022resource}.
Quantum autoencoders have been fruitful in experiments, with several experimental demonstrations~\cite{pepper2019experimental,mok2023rigorous,zhang2022resource}.

Quantum autoencoders have been implemented using variational quantum algorithms~\cite{peruzzo2014variational}, in which the unitary is represented by a parameterized quantum circuit that is trained on a few example states.
To train the circuit parameters, the encoding unitary is learned~\cite{bisio2010optimal,marvian2016universal}. The training can be challenging, requiring one to avoid exponentially small gradients in barren plateaus~\cite{mcclean2018barren}. Further, to successfully converge to a good solution, one often requires
overparameterization~\cite{kiani2020learning,anschuetz2021critical,larocca2021theory,anschuetz2022quantum,you2022convergence,haug2024generalization}.
Additionally, one would like to generalize, i.e. from a few training states drawn from some distribution, the trained circuit is able to perform well on arbitrary states drawn from the distribution~\cite{caro2021generalization,banchi2021generalization,caro2021encoding,caro2022out,gibbs2022dynamical,haug2024generalization}.

However, the literature on quantum autoencoders mainly focuses on reducing the number of qubits, i.e. finding a lower-dimensional representation of quantum states. Compressing information to fewer qubits implies that quantum information is encoded into non-local correlations, strongly increasing the amount of entanglement.
Thus, the current types of quantum autoencoders produce highly entangled states, which are sensitive to noise, and have a complicated representation.

Here, we propose the disentangling quantum autoencoder (DQAE). It transforms entangled $N$-qubit quantum states into a tensor product of $N$ single-qubit quantum states, removing entanglement and factorizing the quantum correlations.
We apply the DQAE for the task of transporting entangled state through a qubit-loss channel. In comparison to unencoded transport, the DQAE has an exponential advantage in the number of copies needed to transfer at least one error-free state. 
The DQAE can be trained in an unsupervised manner using entangled quantum data via variational quantum algorithms. \revA{The corresponding cost function is either based on local Pauli observables or the local purity of the state.}
We show that the gradient can be measured efficiently via Pauli or Bell measurements.
We show that for structured circuit types, barren plateaus can be absent, allowing for efficient training. 
We characterize the generalization capability, i.e. whether training on a few states allows the DQAE to disentangle unseen test states, using the data quantum Fisher information metric (DQFIM). For Ising-type evolution, we astonishingly require only $L=1$ training states and $M\propto N$ circuit parameters to generalize. For stabilizer states, we propose a Metropolis algorithm to train a Clifford ansatz that generalizes using only $L\propto N/4$ training states. 
With our results, quantum information processing with near-term quantum computers can be greatly enhanced.

\section{Model}
\begin{figure}[htbp]
	\centering	
	\subfigimg[width=0.45\textwidth]{}{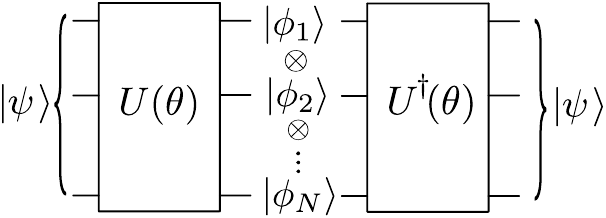}\hfill
	\caption{Disentangling quantum autoencoder (DQAE) disentangles $N$-qubit state $\ket{\psi}$ into a tensor product of $N$ single-qubit states $\ket{\phi_j}$ via unitary $U(\boldsymbol{\theta})$, and can recover the original state again via the inverse $U^\dagger(\boldsymbol{\theta})$. The DQAE can be trained in an unsupervised manner from entangled data. For transferring a quantum state via a qubit-loss channel, DQAE gives an exponential reduction in the number of copies $R$ of the state.
	}
	\label{fig:DQAE}
\end{figure}

The DQAE \revA{unitary} $U$, as shown in Fig.~\ref{fig:DQAE}, disentangles a set $W$ of $N$-qubit quantum states $\ket{\psi}\in W$ into a tensor product of $N$ single-qubit states
\begin{equation}
    U\ket{\psi}=\bigotimes_{j=1}^N \ket{\phi_j}\,,
\end{equation}
where $\ket{\phi_j}$ are single-qubit states that depend on $\ket{\psi}$. \revA{Details of how the unitary $U$ is implemented and trained for such sets are provided in Sec~\ref{sec:training}.}
As $U$ is unitary, one can reconstruct the original entangled state via the inverse $U^\dagger$ 
\begin{equation}
    U^\dagger\bigotimes_{j=1}^N \ket{\phi_j}=\ket{\psi}\,.
\end{equation}
From simple counting arguments, one can see that a single DQAE $U$ cannot compress the set of all possible $N$-qubit states $W=\mathcal{S}(2^N)$ into single-qubit states at the same time. In particular, to represent any $N$-qubit state $\ket{\psi}$ as single-qubit tensor products, one requires $O(2^{N})$ single-qubit states~\cite{plesch2010efficient,rozema2014quantum,yang2016efficient}. 
However, efficient compression is possible when we consider the DQAE for compressing a restricted set $W$ of states.
We note that while we only consider compression to $N$ qubits, one can also reduce the number of single-qubit product states to $K\leq N$. Here, one can further enhance the capacity of the DQAE by including classical information into the $N-K$ qubits that were reduced as shown in Appendix~\ref{sec:capacity}.

\begin{figure*}[htbp]
	\centering	
	\subfigimg[width=0.8\textwidth]{}{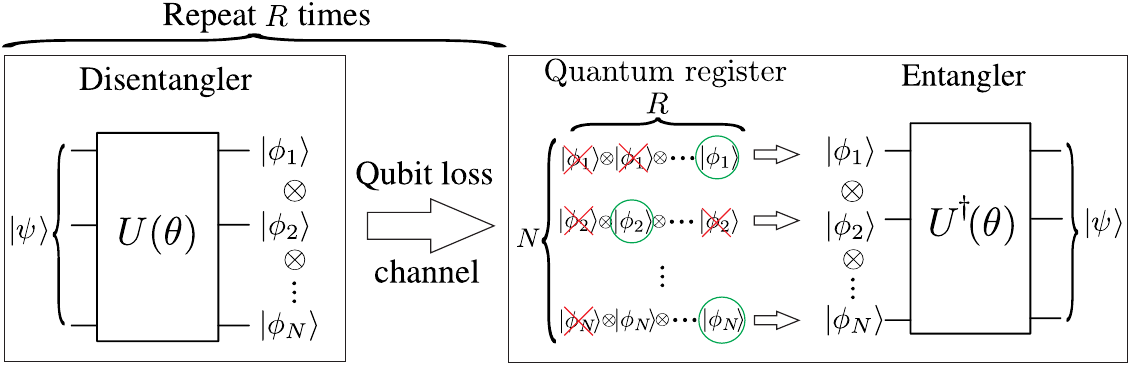}\hfill
	\caption{\revA{\emph{Illustration of the product protocol:}} Disentangling quantum autoencoder (DQAE) to transfer quantum states across qubit loss channels with exponential advantage. Entangled input $N$-qubit state $\ket{\psi}$ is transformed by $U(\boldsymbol{\theta})$ into $\bigotimes_{j=1}^N\ket{\phi_{j}}=U\ket{\psi}$, a tensor product of $N$ single-qubit states $\ket{\phi_{j}}$. The transformed state is then sent through a qubit loss channel $\mathcal{E}$, and stored in a quantum register by the receiver. We repeat this process \revA{for} $R$ copies of $\ket{\psi}$, and finally store $NR$ qubits in the quantum register. Some qubits are lost due to the channel $\mathcal{E}$ (red crosses), which we detected using the leakage detection protocol of Fig.~\ref{fig:Leakage}. At least one of each single-qubit state survives with high probability when $R\propto\log(N)$. In this case, we collect $k=1,\dots, N$ single-qubit states $\ket{\phi_{j}}$ unaffected by loss (green circle). With entangler $U^\dagger$ we then reconstruct the original state $\ket{\psi}=U^\dagger(\boldsymbol{\theta})\bigotimes_{j=1}^N\ket{\phi_{j}}$. 
	}
	\label{fig:sketch}
\end{figure*}

\section{DQAE for qubit-loss channels}
A key application of the DQAE is to deal with qubit loss, which poses a major challenge in quantum computing and communication. Here, we will see that the DQAE can enable an exponential improvement.

\revA{We now propose the DQAE algorithm, which is highly tolerant to qubit loss channels. 
First, let us describe the transport problem that we aim to solve:}
We assume that we have $R$ copies of an $N$-qubit state $\ket{\psi}$. We can send the $R$ copies of $\ket{\psi}$ one by one through a noisy quantum channel $\mathcal{E}(\ket{\psi})$. Now, we want to achieve that at least one copy passes through the channel and arrives as a pure state. %
We assume that each qubit is subject to a qubit loss channel or leakage error~\cite{preskill1998fault}, where the $k$th qubit is transferred to an inaccessible leakage state $\ket{\bot}$ orthogonal to the other qubit states.
\begin{equation}
    \mathcal{E}_k(\rho)=(1-q)\rho+q\text{tr}_{k}(\rho)\otimes\ket{\bot}\bra{\bot}\,,
\end{equation} 
where with probability $q$ it is lost. Here, $\text{tr}_{k}(\rho)$ traces out the $k$th qubit. We assume loss occurs on all qubits equally with the total channel given by
\begin{equation}\label{eq:channel}
\mathcal{E}(\rho)=\mathcal{E}_1\circ\mathcal{E}_2\circ\dots\circ\mathcal{E}_N(\rho)\,.
\end{equation}
For a fully entangled state, loss on any qubit collapses the state into a mixture, and it cannot be used for further processing.
\revA{As a simple example}, let us consider the $N$-qubit GHZ state
\begin{equation}
    \ket{\text{GHZ}}=\frac{1}{\sqrt{2}}(\ket{0\dots0}+\ket{1\dots 1})\,.
\end{equation}
Loss of any qubit collapses the state into an $(N-1)$ qubit mixed state 
\begin{equation}
    \ket{\text{GHZ}}\rightarrow\frac{1}{2}(\ket{0\dots0}\bra{0\dots0}+\ket{1\dots 1}\bra{1\dots 1})\,.
\end{equation}
In contrast, for a single-qubit product state 
\begin{equation}
    \bigotimes_{j=1}^N \ket{\phi_j}\rightarrow\bigotimes_{j\neq k}^N \ket{\phi_j}\,,
\end{equation}
the channel only traces out the $k$th qubit state $\ket{\phi_k}$, while the other $N-1$ qubits remain in a pure state. Thus, product states are more robust, which we will showcase in the following sections.

\subsection{Unencoded transfer}
\revA{Now, let us consider a naive protocol to transport states across the qubit loss channel, which we will see has a low success probability.}
We perform a direct unencoded transfer of the $R$ copies.
Here, we apply channel $\mathcal{E}(\ket{\psi})$ on the pure $N$-qubit entangled state $\ket{\psi}$ and with a chance of $(1-q)^N$ no loss occurs on any qubit. The chance of successfully transferring the state has an exponentially small probability.
Now we repeat the protocol with $R$ copies of $\ket{\psi}$. \revA{We define the \emph{failure probability} $Q_\text{e}$ of the unencoded protocol as the probability that none of the $R$ transmitted copies of the entangled state $\ket{\psi}$ are received intact, i.e., every copy suffers a loss on at least one qubit. Equivalently, the protocol succeeds only if at least one of the $R$ copies survives with no qubit lost.

Hence, the failure probability of the unencoded protocol is}
\begin{equation}
Q_\text{e}=(1-(1-q)^N)^R\,. 
\end{equation}
For \revA{$N\gg-\log(1-q)$}, we find
\begin{equation}\label{eq:copiesEnt}
R\approx\frac{-\log(Q_\text{e})}{(1-q)^N}\,,
\end{equation}
which implies that we require exponentially large $R$ with $N$ for fixed failure probability $Q_\text{e}$.

\begin{algorithm}[H]
\caption{DQAE protocol for qubit-loss channels}
\label{alg:DQAE}
\SetAlgoLined
\LinesNumbered
\KwIn{$R$ copies of entangled $n$-qubit state $\ket{\psi}$, disentangler $U$, qubit-loss channel $\mathcal{E}$}
\KwOut{Recovered state $\ket{\tilde{\psi}}$ or protocol failure}

\For{$i = 1$ \KwTo $R$}{
    \textit{Prepare new copy}: $\ket{\psi}$\;
    \textit{Disentangle}: $\ket{\phi} \gets U \ket{\psi}$\;
    \textit{Transfer qubits}: $\tilde{\rho} \gets \mathcal{E}(\ket{\phi}\bra{\phi})$\;
    Apply leakage detection to each qubit in $\tilde{\rho}$ and keep accepted qubits (Fig.~\ref{fig:Leakage})\; 
}

\uIf{any qubit index $j \in \{1, \dots, N\}$ has no surviving copy across $R$ trials}{
    \textbf{Abort} protocol\;
}
\Else{
    Select one surviving copy for each qubit index $j$ to build product state $\bigotimes_{j=1}^N \ket{\phi_j}$\;
    Reconstruct: $\ket{\tilde{\psi}} \gets U^\dagger \bigotimes_{j=1}^N \ket{\phi_j}$\;
    
}
\end{algorithm}

\subsection{DQAE}
\begin{figure}[htbp]
	\centering	
	\subfigimg[width=0.3\textwidth]{}{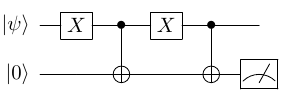}\hfill
	\caption{Leakage detection protocol~\cite{preskill1998fault}. If data qubit $\ket{\psi}$ is present, measurement of the ancilla yields outcome $1$ and leaves $\ket{\psi}$ unchanged. If the data qubit was lost (i.e. became an inaccessible orthogonal state $\ket{\bot}$), measurement of the ancilla yields outcome $0$. 
	}
	\label{fig:Leakage}
\end{figure}

\revA{Now, we show that the DQAE can transport through qubit-loss channels with exponentially fewer copies than the unencoded scenario. Our DQAE algorithm is summarized in Alg.~\ref{alg:DQAE} as well as Fig.~\ref{fig:sketch}. }
\revA{We start by compressing} the incoming $N$-qubit state $\ket{\psi}$ into a single-qubit product state using the DQAE $U$. Here, we assume that we already possess a pre-trained disentangler $U$.
We are now left with an $N$ qubit product state of the form $U\ket{\psi}=\otimes_{j=1}^N\ket{\phi_j}$, where $\ket{\phi_j}\in\mathcal{H}^2$ is a single qubit state. Product states are robust against loss, as in the event of loss, the remaining qubits remain pure.
Our protocol for product states is as follows: We have $R$ copies of $U\ket{\psi}$. After applying $\mathcal{E}(U\ket{\psi})$, some qubits are lost; however, we assume that of the single qubit states $\ket{\phi_j}$, at least one of each has survived.
One can check whether a qubit has survived with the leakage detection protocol~\cite{preskill1998fault} as shown in Fig.\ref{fig:Leakage}, which also has been realized in experiment~\cite{stricker2020experimental}.

We now compute the failure probability of the product protocol. \revA{In this scenario, we define the \emph{failure probability} $Q_\text{p}$ of the product protocol as the probability that for at least one qubit index $j \in \{1,\dots,N\}$, all $R$ copies of $\ket{\phi_j}$ are lost. In other words, the protocol fails unless \emph{each} qubit has at least one surviving copy.} For $N=1$, the probability that at least one qubit survives the channel is $1-q^R$. For $N$ qubits, the product protocol has a failure probability 
\begin{equation}
Q_\text{p}=1-(1-q^R)^N\,.
\end{equation}
Now, we choose $Q_\text{p}\ll1$ and $q^R\ll1$. In this limit, we find the relation
\begin{equation}~\label{eq:copiesProd}
R\approx\frac{\log(\frac{Q_\text{p}}{N})}{\log(q)}
\end{equation}
where $R$ increases only logarithmically with $N$ for fixed $Q_\text{p}$. 

\begin{figure}[htbp]
	\centering	
	\subfigimg[width=0.3\textwidth]{}{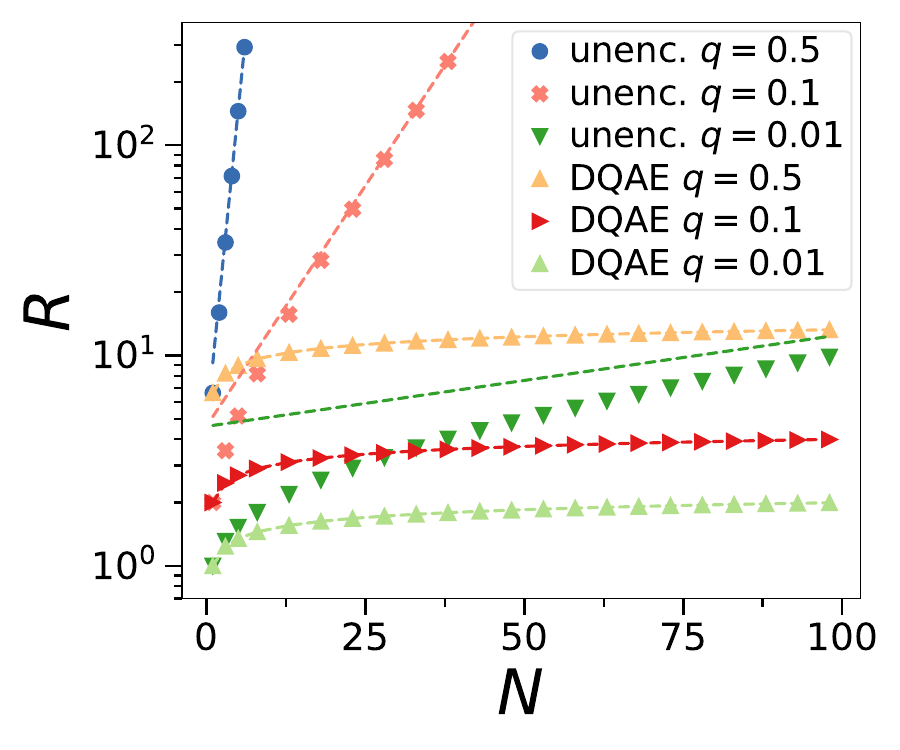}\hfill
	\caption{Number of copies $R$ of an $N$-qubit needed to successfully transfer at least one copy through qubit loss channel of~\eqref{eq:channel}, where each qubit is lost with probability $q$.
    We show transfer with DQAE compared with unencoded transfer of the states for different qubit loss rates $q$. Dashed lines are the approximate formulas~\eqref{eq:copiesEnt} for unencoded and~\eqref{eq:copiesProd} DQAE. We demand that the protocols may only fail with a total failure probability $Q$ of at most $Q\leq 0.01$. %
    }
	\label{fig:copies}
\end{figure}

In Fig.~\ref{fig:copies}, we observe an exponential advantage in copies $R$ for DQAE compared to transferring unencoded entangled states. Further, we find that our formulas~\eqref{eq:copiesProd} and~\eqref{eq:copiesEnt} for $R$ are good approximations of the exact equations.

\subsection{Fidelity}
We now show how to lower bound the fidelity of the protocol of Fig.~\ref{fig:sketch}. We have an entangled state $\ket{\psi}$ and then run our protocol of Fig.~\ref{fig:sketch}.
Let us assume that during the protocol, we send $N$ copies of disentangled $U\ket{\psi}$ through the loss channel. \revA{Further, we assume that for the $k$th copy, we \revA{lose} all qubits except the $k$th qubit, such that the remaining state is given by $\rho_k=\text{tr}_{\bar{k}}(\ket{\psi})$.} 
Here, we define $\text{tr}_{\bar{k}}(.)$ as the partial trace over all qubits except the $k$th qubit. 
We can detect which qubits survived using the leakage detection of Fig.~\ref{fig:Leakage}, and use the reconstruction step to assemble the $N$ remaining qubits $\rho_k$ into a single state $\rho=\bigotimes_k \rho_k$. If the disentangler is optimal, we can perfectly reconstruct the original state via $\ket{\psi}=U^\dagger \rho U$, as each $\rho_k$ is pure.
However, what if the disentangler is not working perfectly? 
\revA{One can quantify the performance of the disentangler using the purity of the reduced density matrix of the $k$th qubit, which is given by
\begin{equation}
E_k(\ket{\psi}=\text{tr}(\rho_k^2)=\text{tr}(\text{tr}_{\bar{k}}(\ket{\psi}\bra{\psi})^2)\,\numberthis\label{eq:purity_loc}\,.
\end{equation}
When purity $E_k=1$, then the $k$th qubit is an unentangled single-qubit state. Thus, an optimal disentangler must have $E_k=1$ for all $k$. }

If the disentangler does not optimally disentangle states, it creates an entangled state with $U\ket{\psi}$ with $E_k<1$ for at least one $k$. 
In this case, for the reconstruction step of Fig.~\ref{fig:sketch}, tracing out for each individual qubit $k$ results in a mixed state $\rho_k=\text{tr}_{\bar{k}}(U\ket{\psi})$. 
The reconstructed state is then given by
$\rho_\text{f}=U^\dagger\bigotimes_{k=1}^N\text{tr}_{\bar{k}}(U\ket{\psi})U$. 
We can compute the fidelity between the original state $\ket{\psi}$ and the reconstructed state as
\begin{equation}
F=\bra{\psi}\rho_\text{f}\ket{\psi}=\bra{\psi}U^\dagger\bigotimes_{k=1}^N(\text{tr}_{\bar{k}}(U\ket{\psi}\bra{\psi}U^\dagger))U\ket{\psi}\,,
\end{equation}
where $F=1$ only when the DQAE encoded state $U\ket{\psi}$ is a product state, and is bounded by $2^{-N}\le F\le1$.

\section{Training}\label{sec:training}
Next, we discuss how to train the DQAE such that it can disentangle a given distribution of quantum states $W$. The training will be performed in an unsupervised manner, i.e. the quantum data has no labels. We sample $L$ training states $\ket{\psi^{(\ell)}}\in W$ as the training set $S_L=\{\ket{\psi^{(\ell)}}\}_{\ell=1}^L$ from $W$. The training states are then used to train the DQAE.
The DQAE is trained by a parameterized unitary $U(\boldsymbol{\theta})$ via $U(\boldsymbol{\theta})\ket{\psi^{(\ell)}}$, where $\boldsymbol{\theta}=(\theta_1,\dots,\theta_M)$ is a $M$-dimensional parameter vector. We minimize a cost function $C_\text{train}(\boldsymbol{\theta})$ by varying the parameter $\boldsymbol{\theta}$. The cost function is given by the average purity of the reduced single-qubit density matrix 
\begin{align*}
C_\text{train}(\boldsymbol{\theta})&=1-\frac{1}{LN}\sum_{\ell=1}^L\sum_{k=1}^N E_k(U(\boldsymbol{\theta})\ket{\psi^{(\ell)}})\,\numberthis\label{eq:cost}.
\end{align*}
We have $C_\text{train}(\boldsymbol{\theta})=0$ only when $U(\boldsymbol{\theta})\ket{\psi^{(\ell)}}$ is a pure single-qubit product state for all $\ell$, else $C_\text{train}(\boldsymbol{\theta})>0$. 

We train the DQAE on the training set of $L$ states and gain the trained unitary $U(\boldsymbol{\theta}^*)$ with parameter $\boldsymbol{\theta}^*$. After training, we evaluate the performance of the DQAE on unseen test data sampled from the same distribution $W$ via the test error
\begin{equation}\label{eq:cost_test}
C_\text{test}(\boldsymbol{\theta})=1-\underset{\ket{\psi}\sim W}{\mathbb{E}}\left[\frac{1}{N}\sum_{k=1}^N E_k(U(\boldsymbol{\theta})\ket{\psi})\right]\,.
\end{equation}
We say that the DQAE generalizes when it achieves low test error, i.e. $C_\text{test}\approx 0$.

We can compute the single-qubit purity needed for the cost function efficiently using two different methods. First, the purity of the reduced density matrix of the $k$th qubit is given by
\begin{equation}
E_k(\ket{\psi})=\text{tr}(\text{tr}_{\bar{k}}(\ket{\psi})^2)=\frac{1}{2}(1+\sum_{\alpha\in\{x,y,z\}}\bra{\psi}\sigma_k^\alpha\ket{\psi}^2)
\end{equation}
where $\sigma_k^\alpha$ with $\alpha\in\{x,y,z\}$ are the Pauli operators acting on the $k$th qubit. The Pauli expectation can be easily estimated on a quantum computer by measuring the respective eigenbasis of the Pauli operators, which requires measuring in $3$ different bases ($x$,$y$, and $z$). 

As an alternative approach, we can use Bell measurements on two copies of the state $\ket{\psi}$. We define the Bell state $\ket{\Psi^-}=\frac{1}{\sqrt{2}}(\ket{01}-\ket{10})$, the $2$-qubit identity matrix $I_2$ and the projector $\Pi_k=I_2^{\otimes k-1}\ket{\Psi^-}\bra{\Psi^-}I_2^{\otimes N-k}$ onto the Bell state of the $k$th qubit pair. It can be shown that~\cite{garcia2013swap,beckey2022multipartite,haug2022scalable}
\begin{equation}\label{eq:SWAP}
E_k(\ket{\psi})=\text{tr}(\text{tr}_{\bar{k}}(\ket{\psi})^2)=1-2\bra{\psi}\bra{\psi}\Pi_k\ket{\psi}\ket{\psi}\,.
\end{equation}
This can be achieved by measuring only in the Bell basis, as the projectors for all $k$ commute. %
\revA{The purity $E_k$  can be estimated with $O(1/\epsilon^2)$ samples within $\epsilon$ additive accuracy via either Pauli or Bell-basis measurements, which can be easily shown using Hoeffding's inequality. As $E_k$ for different $k$ commute, all $E_k$ can be measured in parallel, thus the overall number of samples needed to estimate $C_\text{train}$ is independent of $N$.
}

We now consider a general parameterized quantum circuit of $d$ layers of the form $\ket{\psi(\boldsymbol{\theta})}=\prod_{j=1}^{d} V_j(\theta_j)T_j\ket{0}$ with fixed entangling unitary $T_j$, parameter vector $\boldsymbol{\theta}$ and parameterized rotations $V_j(\theta_j)=e^{-i\theta_j/2\sigma_j}$ generated by some Pauli string $\sigma_j$. For such circuits, gradients $\partial_jE_k(\ket{\psi(\boldsymbol{\theta})})$ can  be measured efficiently with the shift-rule~\cite{mitarai2018quantum}. The shift-rule can be directly adapted to purity measurements~\cite{haug2022scalable}
\begin{align*}\label{eq:grad}
&\partial_jE_k(\ket{\psi(\boldsymbol{\theta})})=\\
&-2(\bra{\psi(\boldsymbol{\theta}+\frac{\pi}{2}e_j)}\bra{\psi(\boldsymbol{\theta})}\Pi_k\ket{\psi(\boldsymbol{\theta}+\frac{\pi}{2}e_j)}\ket{\psi(\boldsymbol{\theta})}\\
&\bra{\psi(\boldsymbol{\theta}-\frac{\pi}{2}e_j)}\bra{\psi(\boldsymbol{\theta})}\Pi_k\ket{\psi(\boldsymbol{\theta}-\frac{\pi}{2}e_j)}\ket{\psi(\boldsymbol{\theta})})\,,
\end{align*}
where $e_j$ is the unit vector for the $j$th entry of the parameter vector $\boldsymbol{\theta}$.

\section{Overparameterization and generalization}
To successfully train the DQAE, we require two conditions: First, training should converge close to a global minimum, i.e. we have a low training error $C_\text{train}$. Additionally, the DQAE needs to generalize from $L$ training data to arbitrary test states, i.e. we have a low test error $C_\text{test}$. 
Training cost is known to converge well when the ansatz circuit is overparameterized, i.e. has enough circuit parameters $M$ to explore the full ansatz space~\cite{you2022convergence,larocca2021theory,haug2021capacity}. This can be evaluated via the data quantum Fisher information metric (DQFIM)~\cite{haug2024generalization}. In particular, the rank of the quantum Fisher information does not increase further when increasing $M$.

Generalization can be similarly evaluated using the DQFIM. An overparameterized model generalizes (assuming no noise) when the rank of the DQFIM does not increase further upon increasing the number $L$ of training states. Note that for $L=1$, the DQFIM is identical to the well-known quantum Fisher information metric~\cite{meyer2021fisher}. 

Let us first define the data state for a given training set $S_L=\{\ket{\psi^{(\ell)}}\}_{\ell=1}^L$~\cite{haug2024generalization} 
\begin{equation}\label{eq:datastate}
\rho_L=\frac{1}{L}\sum_{\ell=1}^{L}\ket{\psi^{(\ell)}}\bra{\psi^{(\ell)}}\,.
\end{equation}
The DQFIM is defined as~\cite{haug2024generalization}
\begin{align*}
&\mathcal{Q}_{nm}(\rho_L,U)=\numberthis\label{eq:dQFIM}\\
&4\mathrm{Re}(\mathrm{tr}(\partial_n U \rho_L \partial_m U^\dagger)-\mathrm{tr}(\partial_n U\rho_L U^\dagger)\mathrm{tr}(U \rho_L\partial_m U^\dagger))\,.
\end{align*}
For a given set of $L$ training states, the maximal rank of the DQFIM is given as~\cite{larocca2021theory,haug2024generalization}
\revA{
\begin{equation}\label{eq:criticalM}
R_L\equiv\max_{M\ge  M_\text{c}(L)}\max_{\boldsymbol{\theta}}\text{rank}(\mathcal{Q}(\rho_L,U(\boldsymbol{\theta}))) \,.
\end{equation}

We say the circuit is overparameterized for a given $L$ when increasing $M$ does not increase $R_L(M)$. The smallest such value defines the critical number of circuit parameters $M_\text{c}(L)$, i.e.,~\cite{larocca2021theory,haug2024generalization}
\begin{equation}
R_L(M) = R_L(M_\text{c}(L)) \quad \text{for all } M \ge M_\text{c}(L)\,.
\end{equation}
}
For overparameterized $M\geq M_\text{c}(L)$, training is likely to converge to the global minima. Note that $M_\text{c}(L)$ depends on the number $L$ of training data, and the circuit parameters needed to overparameterize increase with $L$.

For generalization (i.e. low test error), one has to also supply a critical number of training data $L\ge L_\text{c}$ during training. \revA{The maximal rank is defined as}~\cite{haug2024generalization}
\revA{\begin{equation*}
R_\infty = R_{L_\text{c}}\equiv\max_{L\ge L_\text{c}}R_L(\rho_L,U)\,,
\end{equation*}
The critical number of training states $L_\text{c}$ is given by the smallest $L$ such that the DQFIM rank saturates
\begin{equation}
    R_L = R_\infty \quad \text{for all } L \ge L_\text{c}\,.
\end{equation}
where $L_\text{c}$ is the minimum number of training states that are needed in order to achieve low test error. 
To simplify calculation, it can be approximated by~\cite{haug2024generalization}
\begin{equation}\label{eq:Lcapprox}
    L_\text{c}\approx 2R_\infty/R_1\,.
\end{equation}
Thus, the rank of the DQFIM allows us to determine how many circuit parameters $M_\text{c}$ and training states $L_\text{c}$ are needed to generalize, which we will later apply for the DQAE. 
Note that deep circuits usually concentrate, i.e. random instances of the circuit assume the average value of the ensemble. 
Thus, determining $R_L$ numerically is straightforward without optimization:  One chooses a few random $\theta$ and picks the largest found rank of the DQFIM, which, due to the concentration, is likely to be the maximal rank. 
Note that $\text{rank}(\mathcal{Q}$ and $R_L$ increases monotonously with $M$ and $L$ (in fact usually nearly linear as seen in Fig.2 of Ref.~\cite{haug2024generalization}), thus determining $M_\text{c}(L)$ and $L_\text{c}$ can be done straightforwardly.
In many cases, the rank of the DQFIM is known analytically, such as for hardware-efficient circuits or circuits with a known dynamical Lie algebra~\cite{haug2024generalization}.
}

\begin{figure*}[htbp]
	\centering	
\subfigimg[width=0.99\textwidth]{}{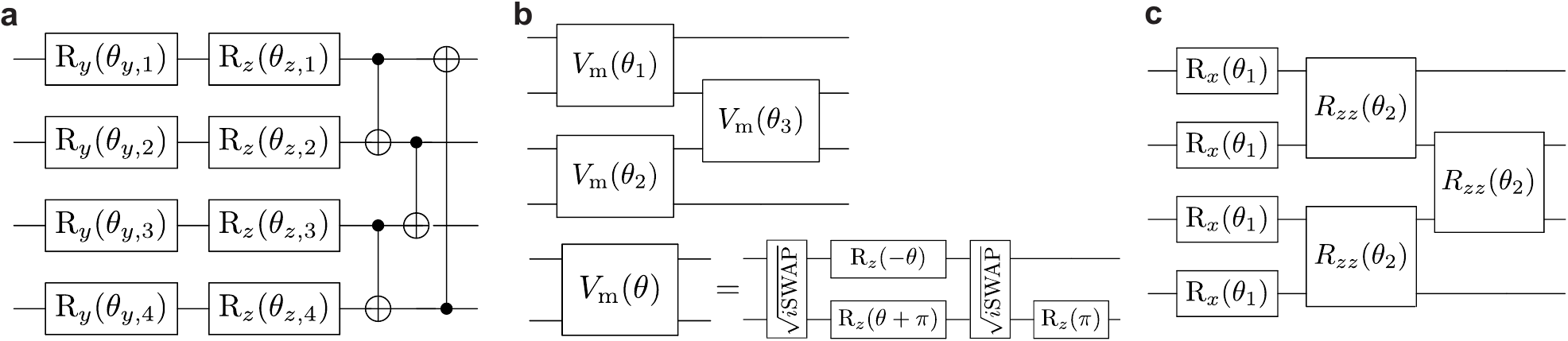}\hfill
	\caption{Parameterized unitaries for disentangler unitary $U(\boldsymbol{\theta})$, which are composed of many layers of the following building blocks: 
 \idg{a} Hardware-efficient ansatz $U_\text{HE}$ of parameterized $y$ and $z$ rotations and fixed CNOT gates arranged in nearest-neighbor fashion.
 \idg{b} Fermionic ansatz $U_\text{fermion}$ 
 \idg{c} Transverse-field Ising inspired ansatz $U_\text{Ising}$.
	}
	\label{fig:ansatze}
\end{figure*}

\section{Ansatz circuits}

We study different types of ansatz circuits as shown in Fig.~\ref{fig:ansatze}. In Fig.~\ref{fig:ansatze}a, we have a hardware-efficient ansatz, constructed from CNOT gates in a 1D nearest-neighbour fashion $W$ combined with layers of parameterized rotations generated from Pauli operators $\sigma^x_j$, $\sigma^z_j$ acting on qubit $j$, which is written as
\begin{equation}
    U_\text{HE}(\boldsymbol{\theta})=\prod_{j=1}^d [\bigotimes_{k=1}^Ne^{-i\sigma^x_k\theta_{2jN+k}}e^{-i\sigma^z_k\theta_{2jN+N+k}}]W\,.
\end{equation}
We also consider a free fermionic ansatz~\cite{google2020hartree}, which is composed of Givens rotations as shown in Fig.~\ref{fig:ansatze}b. 
Additionally, we consider in Fig.~\ref{fig:ansatze}c a circuit that represents the evolution of the Ising model $H=H_{xx}+hH_z$, with $H_{xx}=\sum_i \sigma^x_i\sigma^x_{j+1}$ and $H_z=\sum_i\sigma^z_j$, where we have nearest-neighbor Pauli $x$ interactions, and local $z$ interactions. The evolution under this model of $d$ layers is generated via 
\begin{equation}
    U_\text{Ising}(\boldsymbol{\theta})=\prod_{j=1}^d [\exp(-iH_{xx}\theta_j^x)\exp(-iH_z\theta_j^z)]\,.
\end{equation}

\begin{algorithm}[h]
 \SetAlgoLined
 \LinesNumbered
  \SetKwInOut{Input}{Input}
  \SetKwInOut{Output}{Output}
   \Input{Training data $\{\ket{\psi^{(\ell)}}\}_{\ell=1}^L$

   \hspace{0.2cm}Monotonously decreasing temperature function $T(b)$
   }
    \Output{DQAE $U$}
    $U=I$ 

    $b=0$
    
 \SetKwRepeat{Do}{do}{while}
    \While{$C>0$}{

    $b=b+1$
    
    choose randomly $a\in\{0,1\}$

 \uIf{a=0}{
    random $i\in\{1,\dots,N-1\}$, $j\in\{i+1,\dots,N\}$
    
    $U'=\text{CNOT}_{i,j}U$
  }
  \Else{
      random $i\in\{1,\dots,N\}$
      
      $U_\text{cliff}\in \mathcal{C}_1$
      
    $U'=I_2^{\otimes i-1} U_\text{cliff} I_2^{\otimes N-i}U$
  }
    
    $C'=1-\frac{1}{LN}\sum_{ \ell=1}^L\sum_{k=1}^{N}\text{tr}(\text{tr}_{\bar{k}}(U'\ket{\psi^{\ell}})^2)$

    uniform random $r\in[0,1)$

    $p=\text{min}(1,\exp(-(C'-C)/T(b)))$
    
 \If{$r\le p$}{
    $U=U'$

    $C=C'$
    }

    }
 \caption{Learn Clifford DQAE}
 \label{alg:clifforddisent}
\end{algorithm}

\section{DQAE with stabilizer states}
Stabilizer states and Clifford circuits are one of the fundamental building blocks of quantum information and computing~\cite{gottesman1997stabilizer}.
Due to their simpler structure, they have an efficient algorithm to compute many properties, and even their dynamics can be efficiently learned from quantum states~\cite{leone2022learning}. They are constructed from the Clifford gate set: a combination of CNOT, Hadamard, and S-gates, which can be efficiently simulated classically~\cite{gottesman1997stabilizer}.

Here, we propose Algorithm~\ref{alg:clifforddisent} to learn to disentangle stabilizer states, which is inspired by Ref.~\cite{chamon2014emergent} and the Metropolis algorithm. The goal is to build a disentangling unitary, starting from the identity operator. The idea of the algorithm is to randomly apply either a random single-qubit Clifford unitary $\mathcal{C}_1$ or CNOT gate on randomly chosen qubits. This gate is accepted with probability $p=\text{min}(1,\exp(-(C_\text{p}-C)/T(b)))$, where $T(b)$ is a temperature function which monotonically decreases with each step $b$, $C_\text{p}$ is the cost function with the added gate, and $C$ without. These steps are repeated until the cost function reaches zero.
We empirically find that this algorithm nearly always converges, with only a weak dependence on performance depending on the chosen $T(b)$.

For stabilizer input data, the quantum part of the algorithm can be efficiently simulated on classical computers~\cite{gottesman1998heisenberg}. Note that a classical representation of stabilizer states can be efficiently learned on quantum computers~\cite{montanaro2017learning} using $O(N)$ copies of the state. 
\revA{Thus, we can run our Metropolis algorithm efficiently as follows: First, we learn the stabilizer states of the training set, which can be done using Ref.~\cite{montanaro2017learning} from $O(N)$ measurements. In fact, any stabilizer state can be represented using its generators of the stabilizing Pauli operators, which takes $O(N)$ classical memory. Then, we train the disentangling unitary on a classical computer, where the quantum part is simulated classically using the Tableau formalism for Clifford unitaries~\cite{gottesman1998heisenberg}.}

\section{Results}

We now study the DQAE numerically for different ansatze. 

\subsection{Data generation}\label{sec:datagen}
To generate training and test data, we first draw a fixed unitary $V$ from the respective studied ansatz, as shown in Fig.~\ref{fig:ansatze}.
\revA{Then, for the training data we prepare $L$ random single-qubit product states $\ket{\psi^{(\ell)}}=\bigotimes_j \ket{\phi_j^{(\ell)}}$, which gives us the training set
\begin{equation}\label{eq:trainingset}
    S_L=\{ V\bigotimes_{j=1}^N \ket{\phi_j^{(\ell)}}\}_{\ell=1}^L\,,
\end{equation}
which we use to train our circuit.
The test set is drawn similarly, which is not used during training and only for testing the trained ansatz.
Our choice of $S_L$ ensures that there always exists a DQAE with zero test and training error. However, this solution may be difficult to find during training for the optimization algorithm: For example, convergence to the global minimum is unlikely when the circuit is not overparameterized, i.e. $M<M_\text{c}$. Further, the test error usually remains high when the number of training  data is low, i.e. $L<L_\text{c}$.}

\begin{figure}[htbp]
	\centering	
	\subfigimg[width=0.3\textwidth]{}{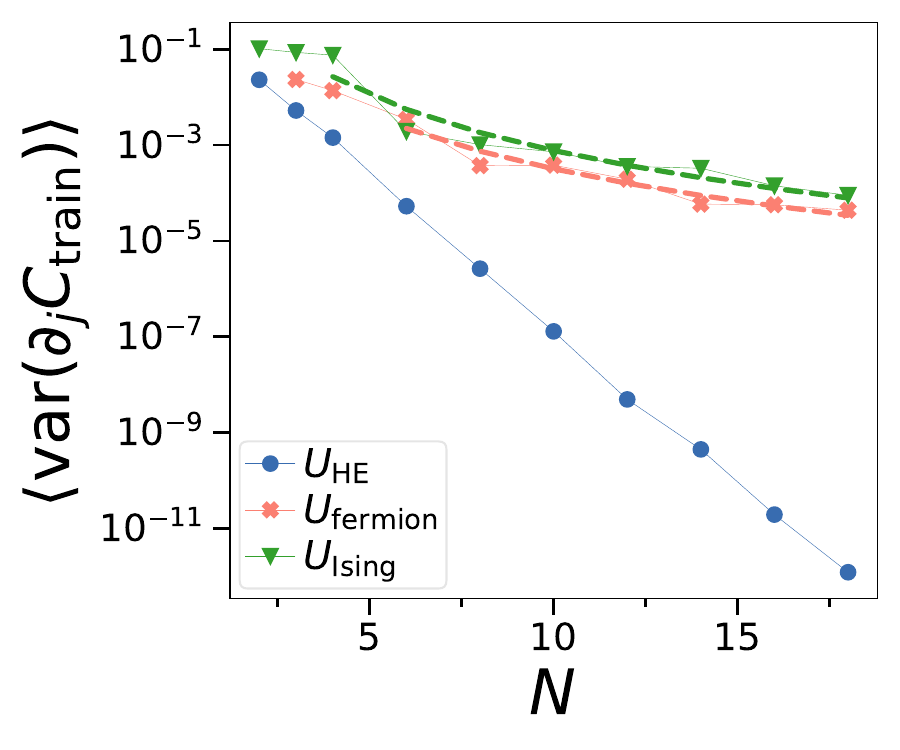}\hfill
	\caption{We show the variance of the gradient for training the DQAE against qubit number $N$. The gradient is computed for random parameters $\boldsymbol{\theta}$ averaged over 10 random instances. We show a \revA{hardware-efficient ansatz} $U_\text{HE}$, fermionic ansatz $U_\text{fermion}$ and Ising model ansatz $U_\text{ising}$. Dashed line is fit with $\langle\text{var}(\partial_j C_\text{train})\rangle=cN^{-\gamma}$, where we find $\gamma=3.80\pm0.46$ for $U_\text{fermion}$, and $\gamma=3.87\pm0.5$ for $U_\text{ising}$. %
	}
	\label{fig:vargard}
\end{figure}
\subsection{Trainability}
Now, we study the trainability of the DQAE, i.e. whether the gradients for training are sufficiently large~\cite{larocca2021theory}.   We show the variance of the gradient $\langle\text{var}(\partial_j C_\text{train})\rangle$ against $N$ in Fig.~\ref{fig:vargard} for different ansatz types as detailed in Fig.~\ref{fig:ansatze}. \revA{Here, we average the variance over all circuit parameters.}
We find that for hardware-efficient ansatze, the variance decays exponentially, which indicates a barren plateau~\cite{mcclean2018barren}. For $U_\text{ising}$ and $U_\text{fermion}$, we find the gradient decays only polynomially, indicating the absence of barren plateaus for these ansatz unitaries, and the circuit can be trained even when the number of qubits is scaled up.

\begin{figure}[htbp]
	\centering	
	\subfigimg[width=0.35\textwidth]{}{trajDN4d50m0t1T100o2000R10D20c0o0n_trainingd21.pdf}\hfill
	\caption{Training error $C_\text{train}$ (solid line) and test error $C_\text{test}$ (dashed) for circuit ansatz $U_\text{HE}$ as shown in Fig.~\ref{fig:ansatze}a. We show the error as a function of the number of training epochs for different numbers of training states $L$ for $M=400$ circuit parameters, where we are in the overparameterized regime for $N=4$. \revA{We have critical number of training states $L_\text{c}=2^N=16$ as derived in Ref.~\cite{haug2024generalization}.}    }
	\label{fig:traj}
\end{figure}

\begin{figure*}[htbp]
	\centering	
    \subfigimg[width=0.9\textwidth]{}{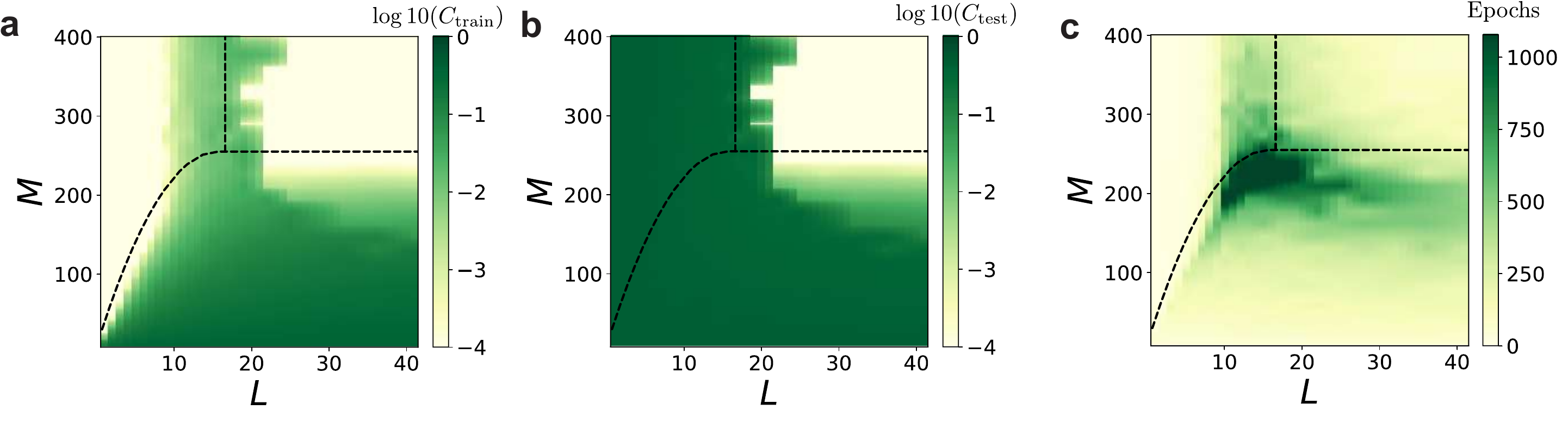}
	\caption{DQAE for disentangling a hardware-efficient ansatz $U_\text{HE}$ as shown in Fig.~\ref{fig:ansatze}a. We show the final \idg{a} logarithm of training loss $C_\text{train}$, \idg{b} logarithm of test loss $C_\text{test}$ and \idg{c} number of epochs needed for training to converge to error $10^{-3}$. We show $N=4$ qubits, for different numbers of training states $L$ and circuit parameters $M$. \revA{Dashed lines are thresholds for overparameterization via~\eqref{eq:criticalM} and generalization via~\eqref{eq:Lcapprox}, which are known analytical as shown in~\eqref{eq:McHardware}. }
	}
	\label{fig:YZ}
\end{figure*}

\subsection{Hardware-efficient ansatz}
Next, we study a hardware-efficient ansatz, shown in Fig.~\ref{fig:ansatze}a, in more detail. 
First, we show a typical training trajectory for the YZ-ansatz in Fig.~\ref{fig:traj} for different numbers of training states $L$. We find three regimes, where we consider the overparameterized limit $M\gg M_\text{c}$: For $L\ll L_\text{c}$, $C_\text{train}$ converges fast, but the test error remains high. For $L\gg L_\text{c}$, dataset is overcomplete and both $C_\text{train}$ and $C_\text{test}$ converge fast to zero. For $L=15\approx L_\text{c}$ close to the critical dataset size, we observe that $C_\text{train}$ takes a long time to converge, with $C_\text{test}$ converging slowly over many epochs.

Next, we study the training and test cost for the hardware-efficient ansatz for different numbers of circuit parameters $M$ and training data size $L$. We find that the training cost in Fig.~\ref{fig:YZ}a becomes small when the circuit is about to become overparameterized, i.e. $M>M_\text{c}(L)$, where the threshold from the DQFIM is indicated as the horizontally aligned dashed line. 
Furthermore, as shown in Fig.~\ref{fig:YZ}b, the test error decreases significantly once the model generalizes, which occurs when the number of training states exceeds $L_\text{c}\approx 2R_\infty/R_1$, as indicated by the vertical dashed line.
We find that around the overparameterization and generalization threshold, the number of epochs needed to train substantially increases, as shown in Fig.~\ref{fig:YZ}c. 
In general, for a hardware-efficient ansatz, both $L_\text{c}$ and $M_\text{c}(L)$ scale exponentially with the number of qubits $N$. \revA{In fact, in this case $R_L$ can be computed analytically, yielding 
\begin{equation}\label{eq:McHardware}
    M_\text{c}(L)=\text{min}(2^{N+1}L-L^2-1,4^N-1)
\end{equation}
and $L_\text{c}=2^N$~\cite{haug2024generalization}.}

\subsection{Ising ansatz}
\begin{figure}[htbp]
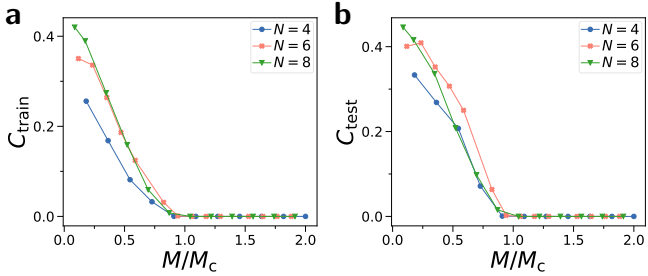

	\centering	
	\subfigimg[width=0.24\textwidth]{a}{costdepthDN4d10m7t2T100o2000R100D20c0o0depthd12.pdf}\hfill
    	\subfigimg[width=0.24\textwidth]{b}{costtestdepthDN4d10m7t2T100o2000R100D20c0o0depthd12.pdf}
	\caption{DQAE for disentangling Ising model evolution $U_\text{ising}$ as shown in Fig.~\ref{fig:ansatze}c. \idg{a} Training and \idg{b} test loss for different qubit number $N$ as function of number of circuit parameters $M$, normalized by critical number of circuit parameters $M_\text{c}$. We have $L=2$ training states for $N=4$, and remarkably $L=1$ for $N=6,8$. 
	}
	\label{fig:ising}
\end{figure}

Next, we study models with polynomial Lie algebra, which can generalize with only polynomial resource cost~\cite{haug2024generalization}.
In particular, in Fig.~\ref{fig:ising} we show the training and test loss for DQAE applied to evolution with the Ising model $U_\text{ising}$ as shown in Fig.~\ref{fig:ansatze}c. 
\revA{Remarkably}, for $N\geq 6$, we require only 
\begin{equation}
    L_\text{c}=1
\end{equation}
training states to disentangle input states. The critical number of circuit parameters $M\geq M_\text{c}$ needed for overparameterization (and thus converging to a global minima) is low, scaling as $M_\text{c}=3N-1$, which we compute via the rank of the DQFIM of~\eqref{eq:criticalM}. 
Thus, for the Ising model, DQAE works remarkably well, requiring only $O(N)$ circuit parameters and $O(1)$ training data.

\subsection{Stabilizer states}
\begin{figure}[htbp]
	\centering	
	\subfigimg[width=0.3\textwidth]{}{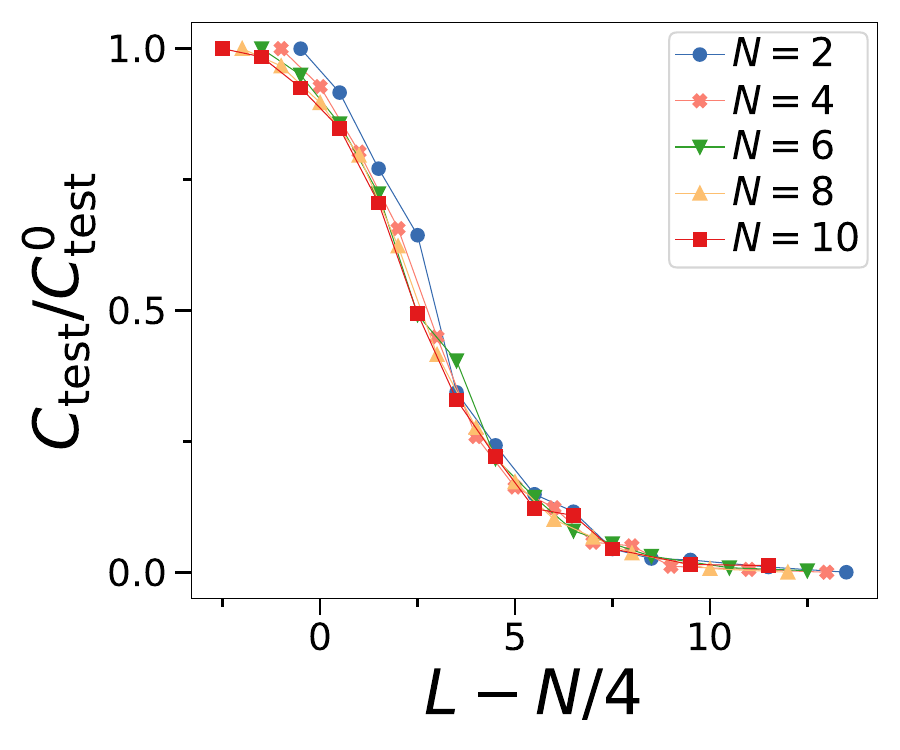}\hfill
	\caption{Test error $C_\text{test}$ for DQAE for stabilizer states as a function of the number of training states $L$. DQAE is trained using the Metropolis algorithm~\ref{alg:clifforddisent}. We plot $C_\text{test}$ normalized to test error without training $C_\text{test}^0$.
	}
	\label{fig:clifford}
\end{figure}

Finally, we apply the DQAE to stabilizer states. We generate training and test data as detailed in Sec.~\ref{sec:datagen}, where we choose $V_\text{Cliff}$ to be a random Clifford circuit, and the input single-qubit states as one of the $6$ single-qubit stabilizer states, i.e. the eigenstates of the $\sigma^x$, $\sigma^y$, and $\sigma^z$ Pauli operators. The DQAE is trained with Algorithm~\ref{alg:clifforddisent}. 
We show the test error for the DQAE after learning stabilizer states in Fig.~\ref{fig:clifford}. We find that the training with the algorithm nearly always converges, and the test error is only limited by the number of training states $L$. We numerically find that for at least up to $N\le10$ qubits, the test errors collapse to a single curve when $C_\text{test}$ is normalized by the test error without training $C_\text{test}^0$ and we shift $L$ by $N/4$. This implies that the number of training states needed to generalize scales as 
\begin{equation}
    L_\text{c}\propto N/4\,.
\end{equation}
Notably, after successfully training the DQAE on stabilizer states, it is also able to disentangle non-stabilizer states of the form  $V_\text{Cliff}\bigotimes_{j} \ket{\phi_j}$, where $\ket{\phi_j}$ are arbitrary (non-stabilizer) single-qubit states. \revA{The fact that for this case generalization can extend beyond stabilizer states has been proven in Ref.~\cite{caro2022out}. Thus, DQAE is capable of out-of-distribution generalization~\cite{caro2022out,haug2024generalization}, i.e. it can generalize even when test and training data are drawn from different distributions.}

\section{Discussion}

We introduced the DQAE to disentangle quantum states into single-qubit product states.
Entangled states are highly susceptible to noise due to the collapse of the wavefunction, while product states are more robust. This allows the DQAE to significantly enhance noise robustness of quantum information processing, such as the transfer of quantum information~\cite{apollaro2022quantum}. 
In particular, for the transfer of entangled states via a qubit loss channel, the DQAE gives an exponential advantage in the number of copies compared to unencoded transfer.

The DQAE can be trained with a simple cost function. 
We characterize overparameterization and generalization using the DQFIM, finding that for product states evolved with an Ising-type evolution, $M=O(N)$ circuit parameters and $L=1$ training data is sufficient for generalization.
We also study DQAE for stabilizer states, finding that $L=O(N)$ training data are sufficient to generalize. Notably, we also showcase out-of-distribution generalization~\cite{caro2022out,haug2024generalization}, demonstrating that training on stabilizer states allows the DQAE to also disentangle a class of non-stabilizer states. 
Our algorithm for disentangling stabilizer states so far rests on a Metropolis-type algorithm. We believe that a direct construction of the disentangling Clifford unitary can be found, similar to other dynamical learning problems for Cliffords~\cite{leone2022learning,yoshida2019disentangling,leone2022retrieving}.

Although we considered the DQAE for a qubit-loss model or (unrecoverable) leakage errors, it can also be applied to quantum error detection~\cite{lidar2013quantum,roffe2019quantum}: In this case, $N$ logical qubits are redundantly encoded into many physical qubits. By performing check measurements, one can detect whether a logical error has occurred and which qubit has been affected.  Let us assume that each logical qubit has a probability $p_\text{L}$ of experiencing an uncorrectable logical error. Without DQAE, one has to discard the full logical state if one logical qubit fails. In contrast, for states disentangled by DQAE, only the affected logical qubit has to be discarded, while the other logical qubits are unaffected. Thus, the DQAE could help to reduce the effective error rate for fault-tolerant quantum information transfer and storing information in quantum memory. \revA{Of course, running the DQAE circuit itself can induce noise, especially on NISQ devices, which have imperfect gates. For the DQAE to effectively reduce noise, we must assume that the (noisy) DQAE circuit itself induces fewer errors than the loss channel. Thus, in practice, one has to consider this tradeoff by taking the actual noisy gate implementation into account. 
} 

\revA{So far, we assumed that qubit loss channels act identically across all qubits. In practice, however, loss rates often vary across different qubits, with loss $q_j$ for the $j$th qubit. Our results still apply when the qubit loss varies: In this case,~\eqref{eq:copiesProd} is an upper bound on $R$ where one chooses $q=\max_j q_j$. }

We note that the disentangling of quantum states could not only be used for noise suppression, but also for learning compact representations of quantum states~\cite{hayashi2023efficient}. Here, the DQAE can serve as a routine in quantum machine learning algorithms to learn a simpler representation of states.

Alternatively, the DQAE could be used to enhance cross-platform verification protocols~\cite{knorzer2022cross}. In particular, measuring the fidelity $\vert\braket{\psi_1}{\psi_2}\vert^2$ between two $N$-qubit quantum states $\ket{\psi_1}$, $\ket{\psi_2}$  on two different devices (without a quantum channel between them) requires exponential complexity $O(2^N)$ in general~\cite{anshu2022distributed}. Instead, one can first apply the DQAE $U$ on both states to disentangle the states into single-qubit product states $U\ket{\psi_1}$, $U\ket{\psi_2}$. Note that this leaves the fidelity invariant. Then, we can efficiently verify the fidelity between the states by local tomography of the $N$ single-qubit states, which scales only as $O(N)$.

Finally, we comment on the relationship of DQAE to the recently proposed Clifford-enhanced matrix product state (CEMPS)~\cite{lami2024quantum} and their disentangling algorithms~\cite{lami2024quantum, huang2024non,fux2024disentangling,liu2024classical,frau2024stabilizer}. Matrix product states (MPSs) are classical state representations with low entanglement, which can be combined with highly entangling Cliffords to be more expressible~\cite{lami2024quantum}. 
Clifford-disentangling of MPS is a classical simulation method that disentangles a (single) quantum state into a CEMPS. Such methods have applications for finding ground states~\cite{qian2024augmenting} and simulating dynamics~\cite{mello2024clifford,qian2024clifford,mello2024hybrid}. Notably, Clifford circuits doped with up to $N$ T-gates can be disentangled into a single-qubit product state with a Clifford unitary applied to it~\cite{huang2024non,fux2024disentangling,liu2024classical}.
We note that, in contrast to Clifford-disentangling of MPS, DQAE is targeted to be trainable on quantum computers and also can disentangle a large class of different input states at the same time.

Future work could implement our DQAE in an experiment on quantum hardware. The DQAE can be straightforwardly implemented, following experimental demonstrations of other types of quantum autoencoders~\cite{pepper2019experimental,zhang2022resource,mok2023rigorous,pivoluska2022implementation}.

\section*{Data availability statement}
The data cannot be made publicly available upon publication because they are not available in a format that is sufficiently accessible or reusable by other researchers. The data that support the findings of this study are available upon reasonable request from the authors. %

\medskip
\begin{acknowledgments}
We thank Paul Bilokon, Michael Hanks and Joschka Roffe for enlightening discussions.
This work is supported by the National Research Foundation of Korea (NRF) grant funded by the Korea government (MSIT) (No. RS-2024-00413957) and the UK Hub in Quantum Computing and Simulation, part of the UK National Quantum Technologies Programme with funding from UKRI EPSRC grant EP/T001062/1. 
\end{acknowledgments}
\bibliography{productencoder}

\appendix 
\onecolumngrid
\newpage 

\section{Increasing model capacity}\label{sec:capacity}
In our discussion of the DQAE, we have so far focused on mapping a dataset of $N$-qubit quantum states to $N$-qubit single-qubit product states. We can generalize this concept to involve a mapping from $N+K$ qubit states to $N$-qubit product states, where $K \geq 0$. The additional $K$ qubits are mapped to a fixed $K$-qubit state $\ket{a}$, referred to as the reference state.

In the original quantum autoencoder architecture proposed by Romero et al.~\cite{romero2017quantum}, the reference state $\ket{a}$ is typically set to $\ket{0}^{\otimes K}$. This state, having no further role in the protocol, is discarded after encoding the quantum state. 

\begin{figure*}[htbp]
    \centering
    \includegraphics[width=0.6\textwidth]{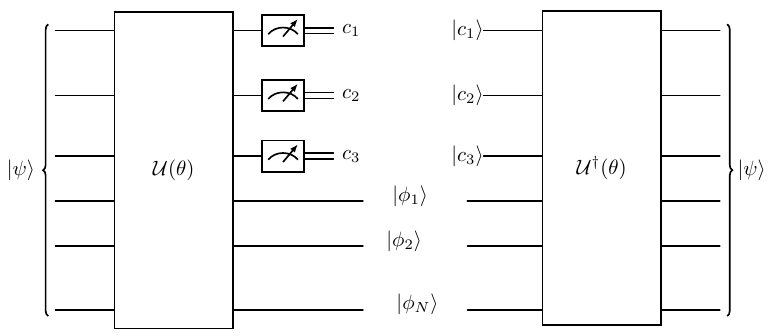}
    \caption{Illustration of an increased capacity DQAE that encodes a $N+K$-qubit state $\ket{\psi}$ into a product state comprising a $N$-qubit single-qubit product quantum state and $K$ classical bits $c_k=\{0,1\}$ by measuring the first $K$ qubits in the computational basis. The entangler $U^\dagger(\theta)$ takes as input computational basis states initialized with the $K$ measurement outcomes, along with the $N$ single-qubit product states, to reconstruct the original state $\ket{\psi}$.}
    \label{fig:increased-capacity-dqae}
\end{figure*}

We propose an alternative approach, where instead of mapping the last $K$ qubits of every state in our dataset to a fixed reference state $\ket{a}$, we allow it to occupy a set of orthonormal states, such as the $K$-qubit computational basis states $\{\ket{i}\}_{i=1}^{2^K}$. 
As we will see, this enhances the capacity of the autoencoder. 
After measurement of the $K$ qubits, the original state can be reconstructed by a modified entangler: One prepares $K$ additional qubits which are initialized with the measurement outcomes $c_k$ of the disentangler. Refer to Fig.~\ref{fig:increased-capacity-dqae} for a sketch of the capacity-enhanced DQAE.

The training of the DQAE can be achieved by slightly modifying the original cost function of the DQAE: We choose $C_\text{train}'=C_\text{train}+\sum_{k=1}^K \bra{\psi}\sigma_{k+N}^z\ket{\psi}^2$, i.e. we add the square of the $\sigma^z_{k+N}$ measurements for each reference qubit. The cost function is minimized when the reference states are one of the computational basis state.

\end{document}